# The effect of biomimetic mineralization of 3D-printed mesoporous bioglass scaffolds on physical properties and in vitro osteogenicity

M. Natividad Gómez-Cerezo[1,2,3], Daniel Lozano[1,2], Daniel Arcos[1,2], Maria Vallet-Regí[1,2], Cedryck Vaquette [3*]

[1]Departamento de Química en Ciencias Farmacéuticas, Facultad de Farmacia, Universidad Complutense de Madrid, Instituto de Investigación Sanitaria Hospital 12 de Octubre i+12, Plaza Ramón y Cajal s/n, 28040 Madrid, Spain

[2]CIBER de Bioingeniería, Biomateriales y Nanomedicina, CIBER-BBN, Madrid, Spain

[3] The University of Queensland, School of Dentistry, Herston, QLD, Australia. *c.vaquette@uq.edu.au

**ABSTRACT**

Three-dimensional Mesoporous bioactive glasses (MBGs) scaffolds has been widely considered for bone regeneration purposes and additive manufacturing enables the fabrication of highly bioactive patient-specific constructs for bone defects. Commonly, this process is performed with the addition of polymeric binders that facilitate the printability of scaffolds. However, these additives cover the MBG particles resulting in the reduction of their osteogenic potential. The present work investigates a simple yet effective phosphate-buffered saline immersion method for achieving polyvinyl alcohol binder removal while enables the maintenance of the mesoporous structure of MBG 3D-printed scaffolds. This resulted in significantly modifying the surface of the scaffold via the spontaneous formation of a biomimetic mineralized layer which positively affected the physical and biological properties of the scaffold. The extensive surface remodeling induced by the deposition of the apatite-like layer lead to a 3-fold increase in surface area, a 5-fold increase in the roughness, and 4-fold increase in the hardness of the PBS-immersed scaffolds when compared to the as-printed counterpart. The biomimetic mineralization also occurred throughout the bulk of the scaffold connecting the MBGs particles and was responsible for the maintenance of structural integrity. In vitro assays using MC3T3-E1 pre-osteoblast like cells demonstrated a significant upregulation of osteogenic-related genes for the scaffolds previously immersed in PBS when compared to the as-printed PVA-containing scaffolds. Although the pre-immersion scaffolds performed equally towards osteogenic cell differentiation, our data suggest that a short immersion in PBS of MBG scaffolds is beneficial for the osteogenic properties and might accelerate bone formation after implantation.

**GRAPHICAL ABSTRACT**

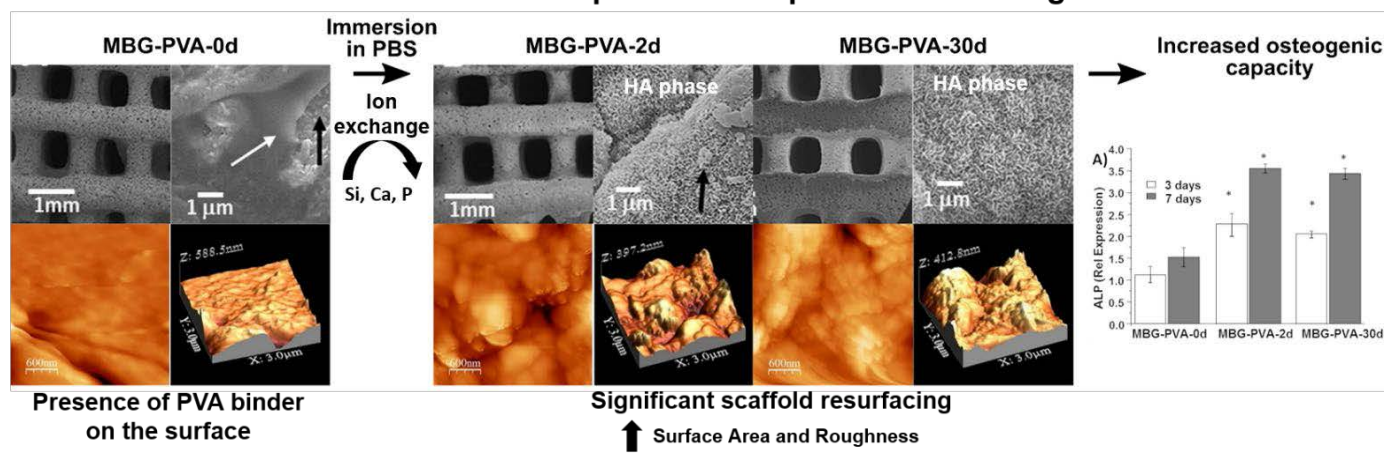

# Introduction

Bioactive glasses have been successfully utilized in orthopedics and dentistry as particulate bone grafting materials due to their bioactivity and space maintenance capability in confined bone defects [1–3]. As an improvement to conventional bioglasses, mesoporous bioactive glasses (MBG) have attracted much attention since their discovery in 2004 [4] due to their excellent textural properties, such as large surface area and porosity, that enhance their bioactive behavior compared to other bioceramics [5]. In fact, MBGs exhibit the fastest bioactive response ever observed for a synthetic material [6]. The high specific surface area of MBGs accelerates ion release in the surrounding microenvironment, initiating the recruitment of progenitor cells and their differentiation towards the osteoblastic lineage, thus enhancing osteoconductivity and osseointegration [7–9].

3D printing technologies has gained significant interest in the last two decades due to their ability to manufacture highly porous bioceramic/polymer constructs suited for bone regeneration [10–13]. Concretely, the use of robocasting for ceramic scaffold manufacturing enables control over the pore size, pore shape and interconnectivity [14] while maintaining some of the bioactivity imparted by the bioceramic particles. Therefore, 3D-printed bioceramic scaffolds combine both the beneficial release of ions from the inorganic material capable of inducing cell differentiation [15–17] and fully interconnected macroporous network, which is a *sine qua non* condition for bone formation, tissue maturation, and hemostasis.

Although the preparation of macroporous scaffolds with mesoporous ceramic materials has been extensively studied, the utilization of additive manufacturing technologies remains challenging due to the extrusion requirements of a colloidal suspension and its subsequent physical stability once extruded/printed. This is usually circumvented by blending the ceramic particles with a polymer binder resulting in a slurry with rheological properties enabling its extrusion through the printing spinneret, hence enhancing the printability of the material [18–20]. However, the presence of a biologically inert polymer significantly decreases the construct's bioactivity and more importantly, can compromise its osteogenic potential, potentially reducing regenerative outcomes [21,22]. For this reason, the polymer is traditionally removed after the printing process in order to obtain a scaffold purely composed of the bioactive ceramic. This is generally achieved by calcination, yet this process induces significant dimensional changes resulting in poor control over the mechanical properties and leads to the modification of the topographical properties of the MBG, hence affecting its bioactivity [23–25]. In order to mitigate the bioactivity reduction inducing the collapsing of the mesoporous structure during calcination [26], a method utilizing a water soluble polymer, polyvinyl alcohol (PVA) has been proposed [20] and offers a number of significant advantages over traditional methods.

PVA is a highly water soluble, biocompatible and biodegradable polymer [27,28], which enables 3D-printing without the use of toxic solvents and does not require removal by calcination. However, PVA often covers a large portion of the scaffold surface, thus hiding the micro- and nano-features of MBGs [25,29]. In turn, this can affect both the scaffold topographical properties (the PVA results in the formation of a smooth surface) and the ions release profile by reducing fluid exchange. Since osteogenic differentiation is directed by topographical cues on the biomaterial's surface, the reduction of textural features of the MBG by PVA can initially alter the scaffold's bioactivity and biological performance. Therefore, removal of the PVA binder and/or the surface modification of MBG scaffolds via the deposition of a biomimetic mineralized layer is sound for the re-establishment of a surface topography and ions exchange suitable for osteogenic differentiation. While numerous studies have reported the capacity of the MBG scaffolds to rapidly undergo in vitro biomineralization (within a time period ranging from 1 to 28 days of immersion) in SBF [18,29,30], little is known on the effect that these surface modifications have towards both physical and biological performances of the resulting scaffolds.

Therefore, this study reports on the physical modifications occurring in PVA-MBG 3D-printed scaffolds during PVA-binder removal in Phosphate Buffered Saline (PBS) (a pH controlled physiological fluid), along with the biological consequences that these extensive surface modifications had on the in vitro scaffold's biological response.

## Materials and methods

**Synthesis of mesoporous bioactive glass.** (MBG). Mesoporous bioactive glass $75SiO_2$-$20CaO$-$5P_2O_5$ was synthesized by evaporated induced self-assembly (EISA) [4]. The slow solvent evaporation in the EISA method allows obtaining highly ordered porous multicomponent structures compared to the conventional sol-gel glasses [30]. Tetraethylorthosilicate (TEOS), triethyl phosphate (TEP), and $Ca(NO_3)_2 \cdot 4H_2O$ were utilized as sources of $SiO_2$, $P_2O_5$, and $CaO$, respectively. Non-ionic triblock copolymer Pluronic F68 (polyoxyethylene–polyoxypropylene–polyoxyethylene type polymer $(PEO)_{75}$-$(PPO)_{30}$-$(PEO)_{75}$) was incorporated as a structure-directing agent. All chemicals were purchased from Sigma Aldrich. Pluronic F68 (4 g) was dissolved in 76 mL of ethanol with 1 mL of 0.5 M HCl solution. Afterward, TEOS (7.6 mL, (0.45 M)), TEP (0.84 mL, (0.065 M)), and $Ca(NO_3)_2 \cdot 4H_2O$ (2.20 g, (0.13 M)) were added under continuous stirring for 3 hour intervals. The solutions were subsequently casted onto Petri dishes to allow the gelation process, which occurred after 35 h, and then gels were aged for 7 days in Petri dishes at 30°C. The dried gels were obtained as homogeneous and transparent membranes that were treated at 700°C for 3 hours in order to obtain the final calcined glass powder. Finally, the materials were gently ground with a glass mortar and subsequently sieved in order to obtain a particle size below 40 µm.

**Additive Manufacturing of MBG scaffolds.** The MBG-PVA scaffolds were prepared via extrusion-based additive manufacturing (AM) method. To this end, 1.5 g of PVA (Mw 31,000-50,000, Sigma Aldrich) was dissolved in 10 mL of water at 80°C. Subsequently, 7.5 g of MBG were gradually added to the solution by stirring in batches of 0.5 g. Thereafter, the mixture was placed in a planetary centrifugal mixer (ARE-250, Thinky Corp, Tokyo, Japan) and mixed at 1800 rpm for 2 min. After these treatments, the mixture exhibited an appropriate viscosity allowing extrusion and 3D printing. Macroporous 3D-printed scaffolds with dimensions of 25 length x 25 width x 3.5 height mm$^3$ were manufactured via the extrusion and solidification of this mixture using a robotic deposition apparatus (EnvisionTEC GmbH PrefactoryVR 3-D BioplotterTM, Gladbeck, Germany) using a 0-90º layout with a 1 mm strut interdistance, a thickness layer of 460 µm, on a collector stage travelling at 300 mm/min. The scaffolds contained 83% of MBG and 17% PVA (wt/wt). Finally, the scaffolds were kept at 30 °C overnight to remove the remaining water and treated at 150 °C for 30 min as was previously reported [31,32].

**Characterization**

**Transmission electron microscopy (TEM).** Transmission electron microscopy (TEM) was carried out using a JEM-1400 microscope (JEOL., Tokyo, Japan), operating at 300 kV (Cs 0.6 mm, resolution 1.7 Å). Images were recorded using a CCD camera (model Keen view, SIS analyses size 1024 X 1024, pixel size 23.5 mm X 23.5 mm) at 50,000x magnification using a low-dose condition.

**Biomineralization and PVA leaching out**. The leaching out of the PVA and the biomineralization of the 3D printed scaffolds was assessed in PBS solution at pH=7.4 and 37 °C for 30 days. Individually tracked scaffolds (n=6) with mass ranging from 55 to 65 mg was immersed in 5 mL of PBS for 2, 7, and 30 days. At each time point, samples were dried at 50 °C for 48 hours and weighted. The entire volume of PBS was replaced at 2, 4, 7, 10, 15, 20, 25 and 30 days. The percentage of mass loss was calculated according to equation 1 below, where $M_1$ is the initial mass, and $M_2$ is the final mass.

$$\text{Mass loss (\%)} = (M_1-M_2)/M_1 \times 100 \qquad (1)$$

The measured mass loss from the scaffold is a combination of three simultaneous and competing phenomena: the removal of the PVA and the ions released from the MBG (negative loss) and the formation of a biomineralization layer on its surface (positive gain).

The scaffold surface modifications occurring during PBS immersion were assessed using various characterization techniques as described below. To this end, the scaffolds were divided into four groups: i) no immersion (MBG-PVA-0d), ii) 2 days of immersion (MBG-PVA-2d), iii) 7 days of immersion (MBG-PVA-7d) and iv) 30 days of immersion (MBG-PVA-30d).

**Scanning electron microscopy (SEM).** Scanning electron microscopy (SEM) and backscattered electrons (BSE) imaging was carried out using a JSM F-7001 microscope (JEOL Ltd., Tokyo, Japan). The scaffolds (n=2) from each immersion time point were mounted onto SEM stubs and carbon coated in vacuum using a sputter coater (Balzers SCD 004, Wiesbaden-Nordenstadt, Germany). Energy-dispersive X-ray spectroscopy (EDS) measurements on each of the two scaffolds were taken for each group. The SEM micrographs were also utilized for determining the average strut dimension (n=2 with 6 measurements in each sample).

**Determination of residual PVA.** In order to determine the residual amount of PVA, thermogravimetric analysis (TGA) was carried out using a TG/DTA Seiko SSC/5200 thermobalance (Seiko Instruments, Chiva, Japan) between 30°C and 650 °C in an air atmosphere at a heating rate of 1 °C·min$^{-1}$, using platinum crucibles and α-Al$_2$O$_3$ as references.

**Specific surface area - Brunauer-Emmett-Teller (BET).** The textural properties of both, MBG particles and, scaffolds, (n=3) were determined by nitrogen adsorption with a Micromeritics ASAP 3020 equipment (Micromeritics Co., Norcross, USA). Prior to the N$_2$ adsorption measurements, the samples were previously degassed under vacuum during 24 h, at 105°C. The surface area was determined using the Brunauer-Emmett-Teller (BET) method [33]. The pore size distribution between 0.5 and 40 nm was determined from the adsorption branch of the isotherm by means of the Kruk-Jaroniec-Sayari Standard (KJS) method [34].

**Ion release study.** The scaffolds (n=6) were immersed for 30 days in 5 mL of PBS at 37°C, and the entire volume of PBS was collected and replaced at 2, 4, 7, 10, 15, 20, 25 and 30 days in order determine the ion release using Atomic Emission Spectroscopy inductively coupled plasma (ICP-OES). The analysis of Ca, Si and P released was quantified through the emission lines 317.93, 213.61, 251.61 nm respectively, on a Varian, model view AX Pro (Varian Inc, Palo Alto, CA, USA). The detected concentrations were in the calibration range between 0.1 and 10 mmol/L. The measurements were performed in the corresponding emission range of equipment (167-758 nm) as previously reported [19,35].

**X-ray diffraction (XRD).** X-ray diffraction was performed using a Philips X'Pert diffractometer equipped with Cu KR radiation (wavelength 1.5406 Å). Low angle XRD pattern were collected in the 2θ° range between 0.6° and 6.5°, with a step size of 0.02° and a counting time of 4 s per step.

High angle XRD patterns were collected in the 2θ° range between 20° and 50°, with a step size of 0.04° and a counting time of 4 s per step (n=3).

**Fourier transform infrared spectroscopy (FTIR).** Fourier transform infrared (FTIR) spectroscopy was carried out in a Nicolet Nexus (Thermo Fisher Scientific) equipped with a Goldengate attenuated total reflectance device (Thermo Electron Scientific Instruments LLC, Madison, WI USA) (n=3).

**Atomic force microscopy (AFM).** Scaffold topography and surface roughness were evaluated using Atomic Force Microscopy (AFM) for each immersion time point. The specimens (n=3 for each group, with 5 measures for each scaffold hence tallying at least 15 measurements per group) were placed in the sample holder disk. Images were acquired in contact mode with a Multimode AFM with Nanoscope IIIa controller using an NP-S cantilever with a spring constant of 0.12 N/m (Veeco Instruments, Santa Barbara, CA).

**Nano-indentation.** The surface hardness was measured using a Hysitron TI 950 Triboindentor (BRUKER) with a conical tip 60º, 5 µm (two different scaffolds per group were utilized with 3 measures per scaffold, n=2 in triplicate).

**Compressive Young's modulus.** Mechanical compression tests were performed using an Instron 5848 microtester with a 500 N load cell (Instron Australia) in PBS at 37 °C. Scaffolds with dimensions of 7 x 6 x 3.5 mm$^3$ (n=5), were subjected to 50% of compression at a rate of 1 mm·min$^{-1}$. The compressive Young's modulus of the scaffolds was calculated using the initial linear portion of the stress versus strain data.

**Cell culture.** An *in vitro* cell culture study was performed to evaluate the effect the scaffold's surface modifications following PBS immersion has on the cells. Three groups were selected and seeded with mice pre-osteoblastic cells (MC3T3-E1) and cultured for up to 7days in either basal (α-MEM medium supplemented with 10% (v/v) of fetal bovine serum (FBS) and 1% (v/v) penicillin/streptomycin) or osteogenic medium (basal medium supplemented with β-glycerolphosphate (50 mg·mL$^{-1}$, Sigma Chemical Company, St. Louis, MO, USA) and L-ascorbic acid (5 mM, Sigma Chemical Company, St. Louis, MO, USA)); i) no immersion (MBG-PVA-0d), ii) 2 days of immersion (MBG-PVA-2d), and iii) 30 days of immersion (MBG-PVA-30d). Prior to cell seeding all scaffolds were sterilized by immersion in 80% ethanol for 15 min and UV-sterilization for 20 min on each surface. Thereafter, the scaffolds were immersed in 1mL of basal medium for 2 hours. The medium was removed and cells were seeded onto the scaffolds at a concentration of 50,000 cells in 20 µL of medium and allowed to attach for 4 hrs. A regular hydration step was performed every 30 min for the first 2 hours and then every hour for another 2 hours. Four hours post-seeding, 1 mL of medium was added in the well and the culture took place over 7 days with a medium change every 2 days.

**Cell metabolic activity.** At day 1, 3 and 7 the metabolic activity of the cells seeded on the various scaffolds (n=6 for each group, each culture condition and each time point) was quantified using Alamarblue® (ThermoFischer) assay. In brief, the scaffolds were moved to new wells and the cells were exposed to resazurin solution for 4 hours at 37°C under $CO_2$ (5%) atmosphere. The fluorescence signal was detected at $\lambda_{em}$ = 590 nm using a $\lambda_{exc}$= 560 nm using a fluorescence spectrometer Biotek Synergy 4 and the percentage of reduction of the resazurin. Thereafter, fresh medium was added to the wells and the culture was continued to the following time point (3 or 7 days).

**ALP activity.** The alkaline phosphatase (ALP) activity of the cells cultured onto the scaffolds after 3 and 7 days was assessed as a marker for osteogenic differentiation. After 3 and 7 days, the scaffolds (n= 5 for each group, each culture condition and each time point) were rinsed with PBS and the cells were lysed in 300 µL of a 0.1% Triton X-100 solution for 1 hr. Thereafter, 50 µL of the ALP lysate was added to 750 µL of p-nitrophenylphosphate solution (10 mmol/L) and incubated for 30 mins at 37°C. The reaction was subsequently stopped by the addition of 50 µL of 1 M NaOH. The absorbance at 410 nm of 100 µL of this solution was measured in triplicate using a Helios Zeta UV–vis spectrophotometer. The results were normalized by the total protein content as measured with a BCA kit (Sigma Aldrich) at 450 nm on the spectrophotometer. The results were expressed as fold increase in the ALP activity, and this was performed by normalizing the MBG-PVA-2d and 30d data by corresponding MBG-PVA-0d values at a given time point (3 and 7 days) for basal and osteogenic media.

**Real time PCR.** The gene expression of MC3T3-E1 cells cultured onto the scaffolds was measured after 3 and 7 days in basal and osteogenic media (n=4 for each group at each time point). Total RNA was isolated from the cells by a standard procedure (Trizol, Invitrogen, Groningen, The Netherlands), and gene expression of various osteoblastic markers was analyzed by real time PCR using an QuantStudio™ 5 Real-Time PCR System (Applied Biosystems, Foster City, CA), as reported previously [29]. Primers for mouse Runx2, osteocalcin (OC), alkaline phosphatase (ALP) was sourced from Assay-by-DesignSM (Applied Biosystems). GAPDH, a housekeeping gene, was amplified in parallel with the tested genes. The relative mRNA expression was calculated as $2^{-\Delta CT}$, where $\Delta CT$ = $CT_{target}$- $CT_{GAPDH}$ Runx2 (Mm00501578 _m1); ALP (Mm00475834_m1; OC (Mm00485009_m1); GAPDH (Mm99999915_g1).

**Cell morphology.** Cells were cultured in basal media for 3 and 7 days on the various 3D scaffolds, which were rinsed twice in PBS and fixed with 3 % glutaraldehyde in PBS for 6 hours. After rinsing 3 times with PBS, the dehydration was carried out in a series of graded ethanol solutions (30, 50, 70, 90 and 95%), with the final dehydration in absolute ethanol 3 times. The samples were dried at vacuum overnight, and finally the scaffolds were mounted onto stubs and gold coated in vacuum using a sputter coater (Balzers SCD 004, Wiesbaden-Nordenstadt, Germany). Scanning electron microscopy was then carried out with a JEOL 7001 scanning electron microscope operating at 15 kV.

**Morphological studies by confocal laser scanning microscopy.** Confocal microscopy was performed to visualize the morphology of the cells attached to the scaffolds cultured in basal media at day 3 and 7. Each scaffold was rinsed twice in PBS and fixed in 4% (w/v) paraformaldehyde in PBS with 1% (w/v) sucrose at 37°C for 2 hours. Thereafter, the scaffolds were rinsed twice with PBS and then the samples were incubated with Atto 565-conjugated phalloidin at a concentration of 0.165 μM (Molecular Probes). The samples were then rinsed with PBS and the cell nuclei were stained with 1 µg /mL 40-6 diamino-20-phenylindole in PBS (DAPI) (Molecular Probes). Confocal microscopy was performed with a confocal laser scanning microscope OLYMPUS FV1200 (OLYMPUS, Tokyo, Japan), using a FLUOR water immersion lens (NA = 1.0). The images were prepared for analysis using Software 3D Imaris to project a single 2D image from the multiple Z sections by using an algorithm that displays the maximum value of the pixel of each Z slice of 1 μm of depth. The resulting projection was then converted to a TIF file using this software. In the images, DAPI and Atto 565–phalloidin were visualized in blue and red, respectively.

**Statistics.** Data is presented as means ± standard deviations and statistical analysis were performed using the Statistical Package for the Social Sciences software (IBM, SPSS Statistic 25).

Since the data set consisted of multiple groups and/or time points, multiple comparisons were made using one way ANOVA followed by a Tukey post hoc test.

## Results and Discussion

### Evolution of surface features

Figure 1A shows the SEM micrograph of MBG-75S particles used to print the scaffolds. As can be observed, the particles presented an irregular morphology with a size below 40 µm as could be expected from the grounding and sieving processes carried out subsequent to MBG synthesis. TEM image and low angle XRD pattern of MBG-75S (Figure 1B and 1C) shows the ordered mesoporous structure of the particles. The XRD pattern shows a maxima that could correspond to the (10) reflexion of a 2D hexagonal p6m structure, which is in agreement with the ordered porosity previously reported for this type of bioceramic [36–38]. In addition, the $N_2$ adsorption experiments confirmed that the pore size distribution has it maxima in 4.49 nm, being the average pore size of MBG-75S.

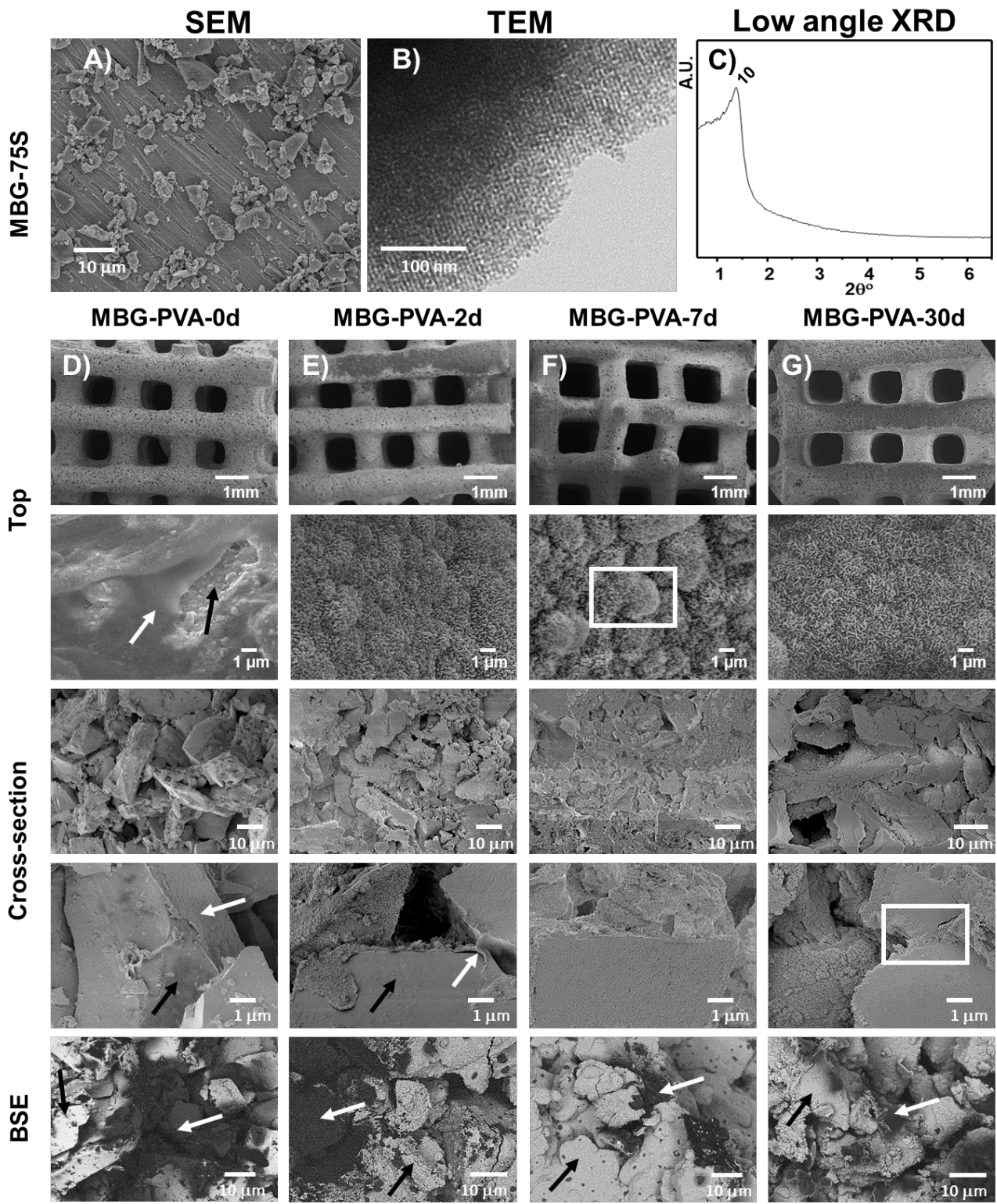

**Figure 1.** Microstructural characterization of MBG-75S particles SEM micrograph A), TEM B), low angle XRD pattern C). SEM and BSE images of the MBG 3D-printed scaffolds before D) and after immersion in PBS for 2 days E), 7 days F) and 30 days G). The black arrows indicate MBG and the new mineralization phase, and the white arrow indicates the presence of PVA. BSE stands for Back Scattered Electron mode.

The 3D-printed scaffolds were immersed for up to 30 days in PBS, and during this period, they retained their physical integrity and maintained their macroporous structure. Figure 1 shows the SEM micrographs of the 3D printed scaffolds at various PBS immersion times. The internal scaffold organization consisted of alternating layers of struts with a 0/90° pattern with a 690 ± 33 µm interstitial distance, which slightly increased throughout the PBS immersion, although it did not reach statistical significance (Table 1). The preservation of the scaffold architecture is of paramount importance as interconnected macroporosity allows angiogenesis [3,39] and supports cell migration and proliferation within the most central portions of the scaffold [40,41].

Table 1. Interstitial distance between the scaffold struts after PBS immersion at various timepoints.

| Sample | Interstitial distance (mm) |
|---|---|
| MBG-PVA-0d | 0.69 ±0.03 |
| MBG-PVA-2d | 0.70±0.03 |
| MBG-PVA-7d | 0.78±0.06 |
| MBG-PVA-30d | 0.76±0.05 |

In addition to the macroscopic characteristics, chemical composition, surface chemistry and topography are also responsible for cellular differentiation [42,43]. The MBG-PVA-0d scaffolds presented a dual topography composed of both smooth and granular surfaces due to the presence of the PVA and the MBG particles, respectively (Figure 1 white and black arrows respectively). The presence of only 17% of PVA resulted in covering most of the surface of the scaffold and significantly obstructed the pores of the MBG particles. Although rarely reported in the literature, the mixing of polymer and bioceramic particles usually results in the formation of a polymer skin over the inorganic materials which impedes cells/bioceramic interaction. In contrast, the immersion of the MBGs scaffold in PBS caused significant changes on the surface of the scaffolds, via the formation of an apatite phase, which nucleated and grew as early as 2 days post-immersion in PBS.

The scaffold topography was modified by the deposition of this biomineralized coating, which presented all the topographical characteristics of an apatite-like phase [44,45]. In addition, the presence of the PVA on the surface of the scaffolds struts was no longer observed at this early immersion time point (Figure 1C-G). After 7 days in PBS, the mineralized layer grew thicker resulting in the formation of micro-scaled nodules (Figure 1 F, square arrowhead). These microfeatures disappeared after 30 days of immersion, leaving a rough but relatively homogeneous surface at the microscale (Figure 1 G).

While the presence of PVA was no longer observed on the scaffold surface after two days post-immersion in PBS, the BSE studies on the scaffold cross-sectional area demonstrated the maintenance of the PVA in the most central portions of the printed struts even after 30 days of immersion. BSE studies shows the presence of the polymer in the cross-section, being the PVA darker than the ceramic in the image. Interestingly, the biomineralization layer found on the scaffold surface was also present within the core of the scaffold as early as 2 days post-immersion and grew around the MBG particles. For longer immersion periods, the biomimetic layer forming around the MBG particles appeared thicker and bridged adjacent paticles (Figure 1 G white rectangle), which contributed to the maintenance of the structural integrity and mechanical properties of the 3D-printed scaffold. Therefore, the immersion of the 3D-printed MBG scaffolds in PBS resulted in the partial replacement of the PVA binder with a biomimetic mineralized layer acting like a cement between the MBG particles.

The presence of PVA in the scaffolds before and after the soaking in PBS was quantified by TGA (Table 2). The thermogram revealed that 7% of residual PVA remained in the scaffold after 30 days in PBS. Initially, a rapid removal within the first 2 days was observed whereby 5% of the PVA had leached out. Afterwards, the rate of PVA removal decreased and an additional 5% was removed, albeit over an extended period of time (28 days). Similarly, the mass loss analysis revealed a rapid initial decrease of 6% within the first 2 days. Thereafter the rate of mass loss decreased and the mass loss reached 11.8% at 7 days, which corroborated previous findings [46]. After 30 days post immersion the scaffolds lost 14% of its original mass despite the thickening of the apatite-like layer.

Table 2. Mass loss and residual PVA content in MBG-PVA scaffolds before and after immersion in PBS (n=5).

| Nomenclature | Time in PBS (days) | Mass loss (%) | TGA Residual PVA (%) |
|---|---|---|---|
| MBG-PVA-0d | 0 | - | 17 ± 0.3 |
| MBG-PVA-2d | 2 | 6.0 ± 0.6 | 12 ± 0.8 |
| MBG-PVA-7d | 7 | 11.8 ± 0.9 | 10 ± 0.1 |
| MBG-PVA-30d | 30 | 14.4 ± 3.4 | 7 ± 0.9 |

All scaffolds demonstrated significant mass loss despite the formation of an apatite layer on both the surface and in the core of the scaffold, suggesting that amongst all the competing phenomena (removal of PVA, ions release from the scaffold and the formation of the apatite layer on the surface), the PVA leaching was the most critical parameter affecting the scaffold mass. The remaining PVA was mostly located in the inner most locations of the scaffolds, where its dissolution was prevented by decreased water diffusion.

We further investigated the influence of the mineralized layer upon the specific surface area as measured by nitrogen adsorption (Table 3). The results demonstrated that the initial removal of the PVA from the surface of the struts and the simultaneous biomineralization resulted in a 3-fold increase in the specific surface area after only 2 days post-immersion. This suggested that both the removal of the superficial PVA, resulting in exposing the MBG particles, and the deposition of the mineralized layer on the scaffold surface was responsible for this increase. The BET surface area remained constant at 7 days post-immersion and subsequently dropped after 30 days, probably due to the late re-surfacing observed in the MBG-PVA-30d samples, which resulted in the thickening of the CaP layer leading to the formation of a more homogeneous surface. For comparison puposes, the specific surface area of the MBG particles (not-printed and in a powder form) was measured and it revealed that it was significantly higher than any of the 3D-printed scaffolds including those previously immersed in PBS (Table 3). This indicates that the 3D-printing process and the presence of PVA or its residual amount after immersion was still significantly affecting this parameter. The impact of the biomimetic apatite formation on the surface properties was also evaluated by AFM, which showed a significant increase in the surface roughness (Figure 2) with increasing immersion time, ultimately confirming SEM and BET findings. Indeed, the MBG-PVA-0d featured a very smooth topography with a roughness of 9.5 ± 3.1 nm, whereas all other post-immersion groups displayed a rougher topography consisting of peaks and valleys in the hundredth of nanometre range with a roughness of 40-50 nm. Although the MBG-PVA-2d roughness was higher (56.7 ± 13.9 nm) than the MBG-PVA-7d and MBG-PVA-30d (48 ± 9.4 nm, and 38.5 ± 15.4 nm respectively), it did not reach statistical significance.

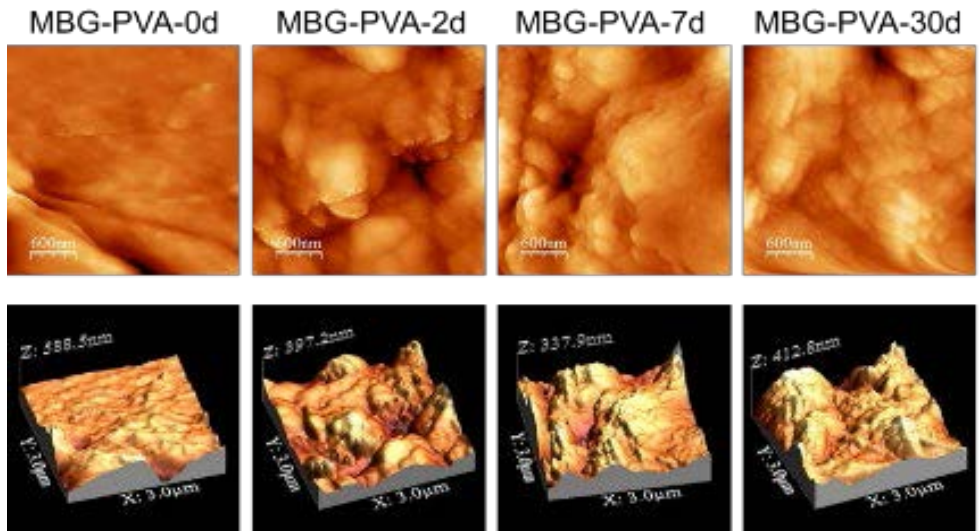

**Figure 2.** Representative AFM images of the scaffold surface after immersion in PBS for 2, 7 and 30 days demonstrating significant changes in the roughness.

Table 3. Textural properties and roughness values of the MBG-PVA 3D-printed scaffold at various PBS immersion time points. The * and # represent statistical difference to all other groups ($p < 0.05$).

| Sample | BET surface ($m^2/g$) | Pore size (nm) | Pore volume ($cm^3/g$) | Roughness (nm) |
|---|---|---|---|---|
| MBG-PVA-0d | 58.3 ± 6.3* | 5.92 ± 0.19 | 0.066 ± 0.005 | 9.48 ± 3.10* |
| MBG-PVA-2d | 186.9 ± 5.9 | 6.88 ± 0.01 | 0.34 ± 0.110 | 56.74 ± 13.86 |
| MBG-PVA-7d | 171.9 ± 10.4 | 8.47 ± 0.01 | 0.33 ± 0.014 | 48.04 ± 9.43 |
| MBG-PVA-30d | 137.7 ± 12.2 # | 9.57 ± 0.48 | 0.23 ± 0.023 | 38.53 ± 15.39 |
| MBG particles | 271.3 ± 13.1 * | 4.49 ± 0.25 | 0.27±0.04 | - |

As seen above, the immersion of the MBG scaffolds in PBS, leading to the deposition of a biomineralized layer and the removal of the PVA, drastically affected the surface properties of the construct. In addition, the presence of the mineralized layer also affected the release of ions contained in the MBG particles, the rate of Si release was decreased as the layer became thicker overtime (Figure 3A). The release of calcium and phosphorus followed an intricate pattern, whereby release and re-precipitation in the mineralized layer were the two main competitive phenomena. Indeed, a small quantity of $Ca^{2+}$ was released initially during the first days of immersion, and subsequently, the calcium release dropped until day 7. From 7 to 15 days post-immersion, there was a slight increase in the amount of $Ca^{2+}$ released, and after 15 days, the $Ca^{2+}$ concentrations continued oscillating around 1 ppm up until 30 days post-immersion. The phosphorus release displayed a similar pattern, that is, a sharp decrease in the first 2 days followed by a gradual return to the PBS initial concentration from day 7 post-immersion. This pattern, whereby the majority of the ions are consumed within the first 7 days, coincided with the early formation and growth of the biomineralized layer as seen by SEM. Subsequently, the biomineralized layer underwent remodelling and therefore did not consume a large quantity of ions, which mirrored the calcium and phosphorus release profile from day 15 to day 30. Overall, the $Ca^{2+}$ release was surprisingly low, and the phosphorus release pattern was not in accordance with previous reports for a similar grade of MBG scaffolds, although immersed in a different release medium, typically Simulated Body Fluid (SBF) [19,36,47]. This discrepancy can be accounted for due to the difference in ion concentrations between SBF, a solution rich in calcium and phosphorus, and PBS with a minimum concentration of calcium (~0.2 ppm) and an elevated amount of phosphorus (~240 ppm) [48]. The deficiency in calcium of the PBS solution resulted in the immediate and almost entire utilization of the calcium released from the MBG scaffolds for the formation of the biomineralization layer (probably, during the first hours of immersion) explaining its low release. As the apatite layer was continuously growing on the surface and in the core of the scaffold, the calcium released remained low throughout the immersion in PBS, even during the resurfacing events previously described.

**Figure 3.** Ion release profile as measured by Atomic Emission Spectroscopy inductively coupled plasma (ICP-OES). A: Silicon, B: Phosphorus and C: Calcium.

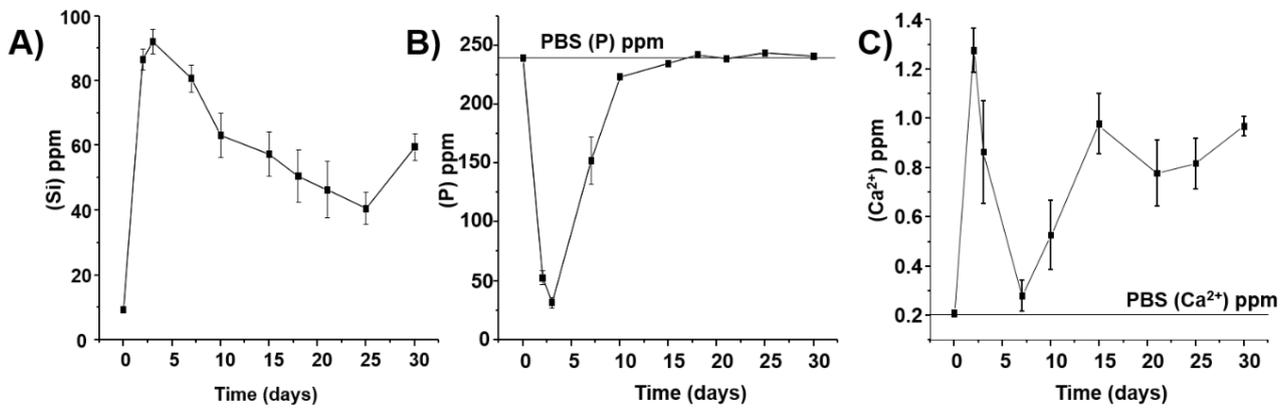

Chemical microanalysis carried out by EDX spectroscopy (table 4) demonstrated that the atomic Ca/P ratio, initially at 3.39 ± 0.43, decreased to 2.48 ± 0.68 at 7 days and subsequently dropped to 1.77 ± 0.57 after 30 days post-immersion, a value close to the Ca/P ratio of hydroxyapatite [49]. This confirmed at a compositional level the extensive remodeling processes experienced by the mineralization layer during the immersion in PBS.

Table 4. Atomic Ca/P ratio calculated from the EDS analysis. The star represent statistical ($p < 0.05$) to all other groups.

| Sample | Ca/P |
|---|---|
| MBG-PVA-0d | 3.39±0.43* |
| MBG-PVA-2d | 2.21±0.18 |
| MBG-PVA-7d | 2.48±0.68 |
| MBG-PVA-30d | 1.77±0.57 |

We further investigated the effect of the remodeling on the biomineralized layer phase using FTIR and XRD analysis (Figure 4). The FTIR spectra for MBG-75S and MBG-PVA-0d (Figure 4A) showed a predominant (Si-O) band at 1100 cm-1. However, a doublet (P-O) at 580 cm$^{-1}$, characteristic of a hydroxyapatite crystalline-like phase, appeared 2 days post-immersion and became gradually more intense and sharp, indicating the presence and subsequent growth of a crystalline calcium phosphate phase. The XRD analysis confirmed these findings and demonstrated a gradual change from an amorphous phase to crystalline hydroxyapatite during PBS immersion. The prolonged immersion in PBS increased the apatite crystallinity, as can be observed by the higher definition of diffraction maxima in the XRD pattern (Figure 4B). Indeed, the diffraction maxima at 25.2, 31.8, 39.8 and 45.5, 2θ°, assigned to the P63/m space group of hydroxyapatite, were observed for the MBG-PVA-7d and MBG-PVA-30d scaffolds (albeit more pronounced for the MBG-PVA-30d) whereas the MBG-PVA-2d scaffolds presented an incomplete pattern (Figure 4B).

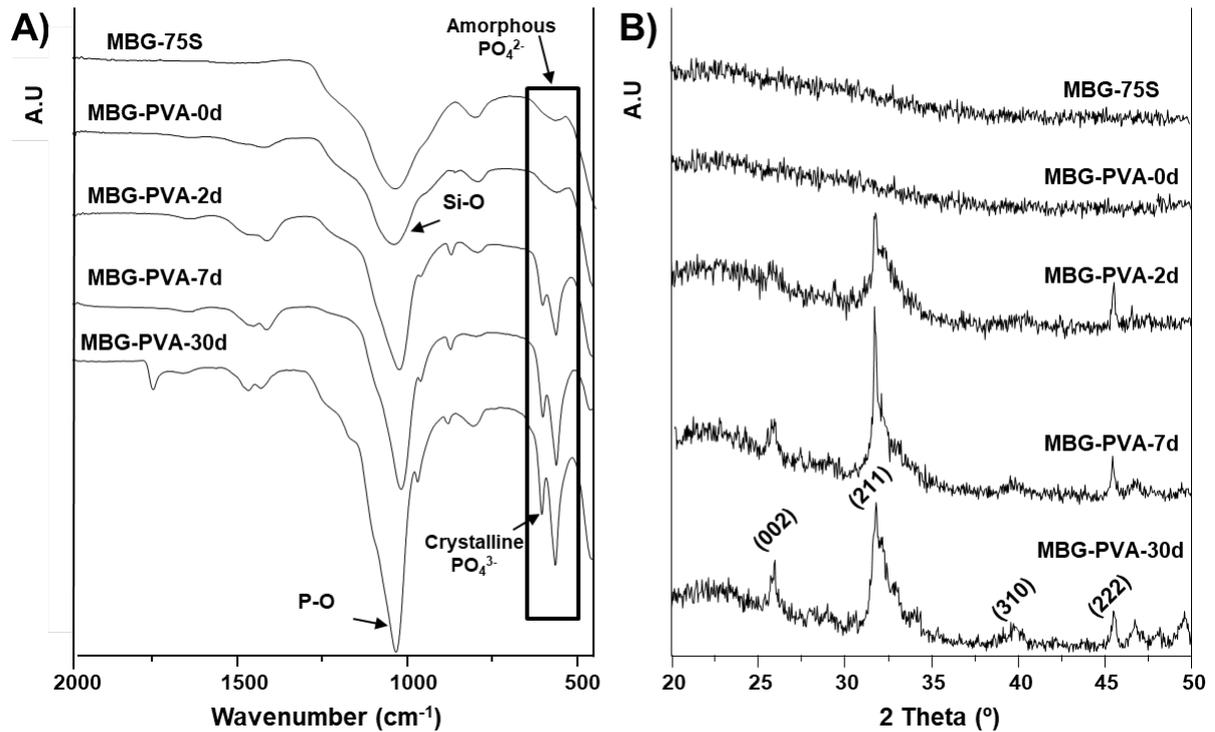

**Figure 4.** Surface characterization of the MBG after calcination and the scaffolds before and after immersion in PBS A) FTIR and B) XRD. Miller indexes for a hydroxyapatite phase are shown between brackets.

**Mechanical properties**

The mechanical properties of the scaffolds were assessed at both the micro- and macroscopic levels using nano-indentation and compression testing. Our results indicate that the deposition of a mineralized layer over the scaffold also caused a drastic change in the surface hardness, as shown in Figure 5A, which revealed a 3-fold increase in this parameter as early as 2 days post-immersion. Despite a slight increase after 30 days of immersion, thereafter, the hardness remained constant (no statistical differences) regardless of the re-surfacing events and phase composition changes previously observed by SEM and XRD, respectively. Figure 5B shows representative compression curves of the MBG scaffold at 0, 2, 7, and 30 days post-immersion in PBS. The scaffolds behaved as typical porous ceramics, displaying a sharp increase in the load followed by failure at low strain (typically around 10%) [50]. The compressive modulus remained relatively constant throughout the immersion in PBS, even though a slight but non-significant, decrease was observed for the MBG-PVA-30d group (Figure 5C). The compressive moduli determined for these scaffolds were lower than what previous reported for similar scaffolds [20]. The compression testing conditions can be accounted for the reduction in the measured modulus as most reports have performed compression test in dry state at room temperature whereas we performed the tests in PBS at 37°C. The strain at failure was in the range of 10% and a small but significant increase was observed for the PBS-immersed groups (Figure 5D). The maintenance of the compressive modulus was attributed to the formation of the calcium phosphate layer on the scaffold surface and around the MBGs particles as observed by SEM (Figure 1 G, cross-section, white box). Indeed, it was hypothesized that the apatite layer acted as a strong binder of the MBG particles enabling excellent maintenance of the scaffold structural integrity and mechanical properties. Although the effect of biomineralization on mechanical properties has been reported previously for individual MBG particles [51], it had not been described for a 3D-printed MBG porous scaffold.

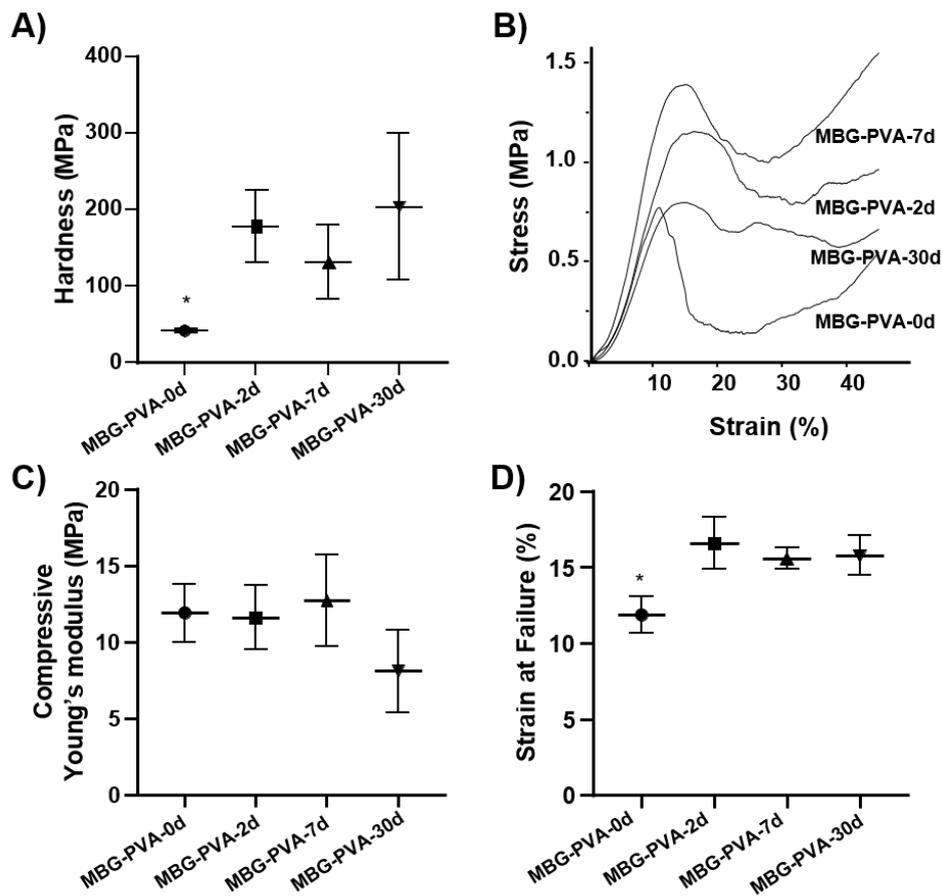

**Figure 5.** Mechanical properties of the scaffolds during the treatment in PBS A) Hardness, B) stress-strain curves, C) compressible Young's modulus and D) Strain at failure. The star represent statistical ($p < 0.05$) to all other groups.

**Cell culture**

In order to investigate the influence of topographical changes on cell proliferation and differentiation, murine pre-osteoblastic cells (M3CT3-E1) were seeded onto the scaffolds from days 0, 2 and 30 post-immersion. These groups were selected as they exhibited the most striking topographical differences compared to the as-printed scaffold. The results showed that the resurfacing of the scaffolds occurring during the PBS immersion had a significant impact on cell metabolic activity (Figure 6A). Indeed, cells cultured in basal media on the scaffolds previously immersed in PBS had significantly higher metabolic activities compared to the as-printed scaffold (MBG-PVA-0d) at both 3 and 7 days post-seeding. However, this trend was not observed when the cells were cultured in osteogenic medium as all groups behaved in a similar manner (Figure 6B).

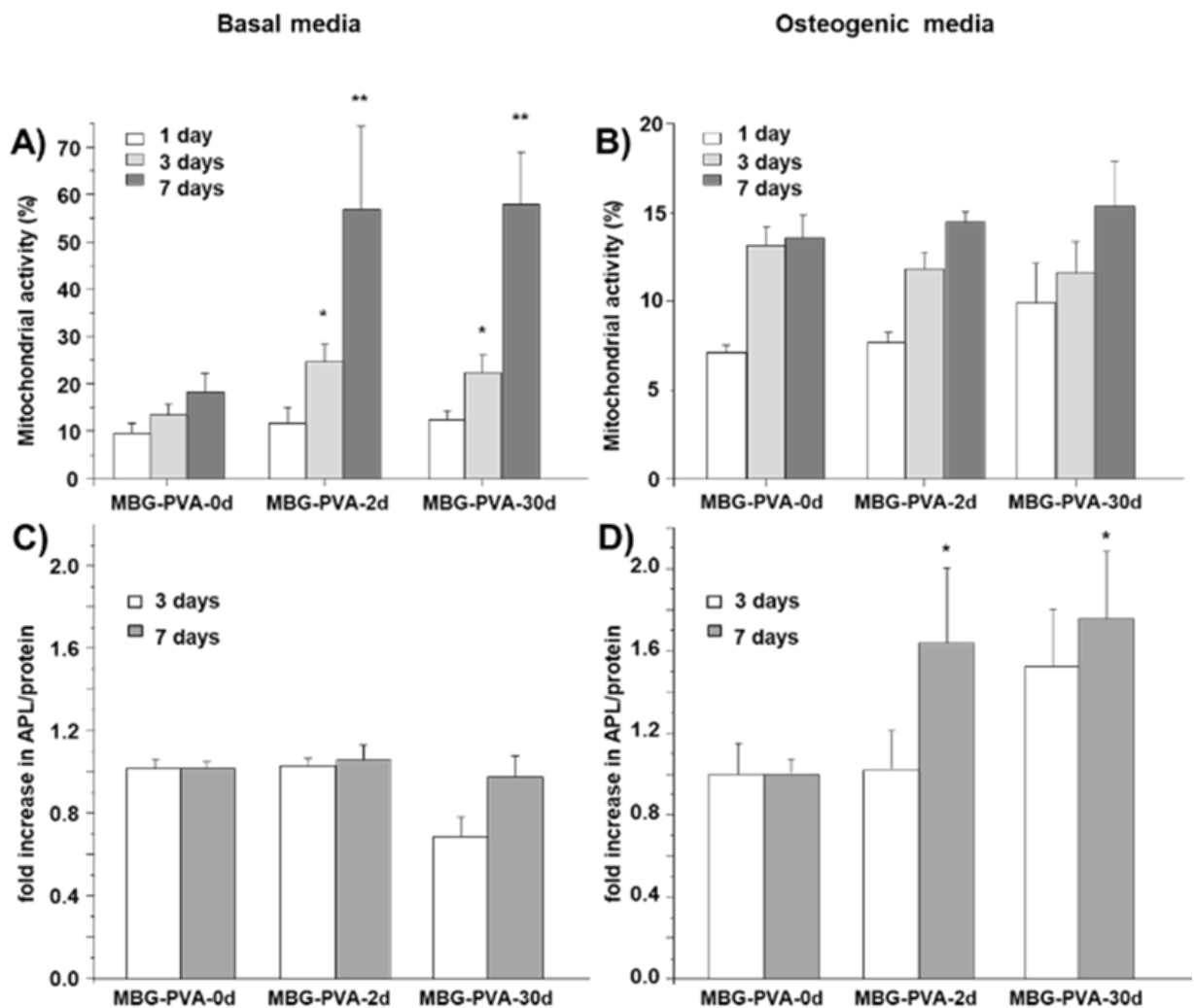

**Figure 6.** Cell Metabolic activity of MC3T3-E1 cells at 1, 3 and 7 days A) in basal media, B) in osteogenic media. Fold increase in ALP activity at 3 and 7 days C) in basal media and D) in osteogenic media (**<0.01 and *<0.05 show statistical significance compared to the normalized MBG-PVA-0d at the corresponding timepoint)

ALP activity was utilized as a marker of osteogenic differentiation for M3CT3-E1 pre-osteoblast like cells. The changes observed in the surface of the scaffolds did not have a substantial impact on early ALP activity, when the cells were cultured in basal media (Figure 6C). However, the combination of the scaffold resurfacing and osteogenic medium acted in a synergetic manner leading to a significant increase in the ALP production at 7 days post seeding for the MBG-PVA-2d and MBG-PVA-30d scaffolds when compared to the as-printed scaffold MBG-PVA-0d (Figure 6D). In order to assess the cell commitment towards osteogenic differentiation, gene expression of bone-related markers (ALP, Osteocalcin and Runx2) was measured by qPCR. The results revealed significant differences in the gene expression pattern depending on the scaffolds surface. Indeed, both MBG-PVA-2d and MBG-PVA-30d scaffolds showed significantly increased gene expression of all of the bone-related markers at 3 and 7 days post-seeding, in both basal and osteogenic conditions (Figure 7) when compared to the MBG-PVA-0d scaffold at the corresponding time point. Furthermore, the gene expressions of these markers were higher in osteogenic condition compared to basal media, as expected. The most significant differences observed between the groups was seen at 7 days post-seeding as all the genes displayed a 2-fold increase for both MBG-PVA-2d and MBG-PVA-30d when compared to the as-printed scaffolds.

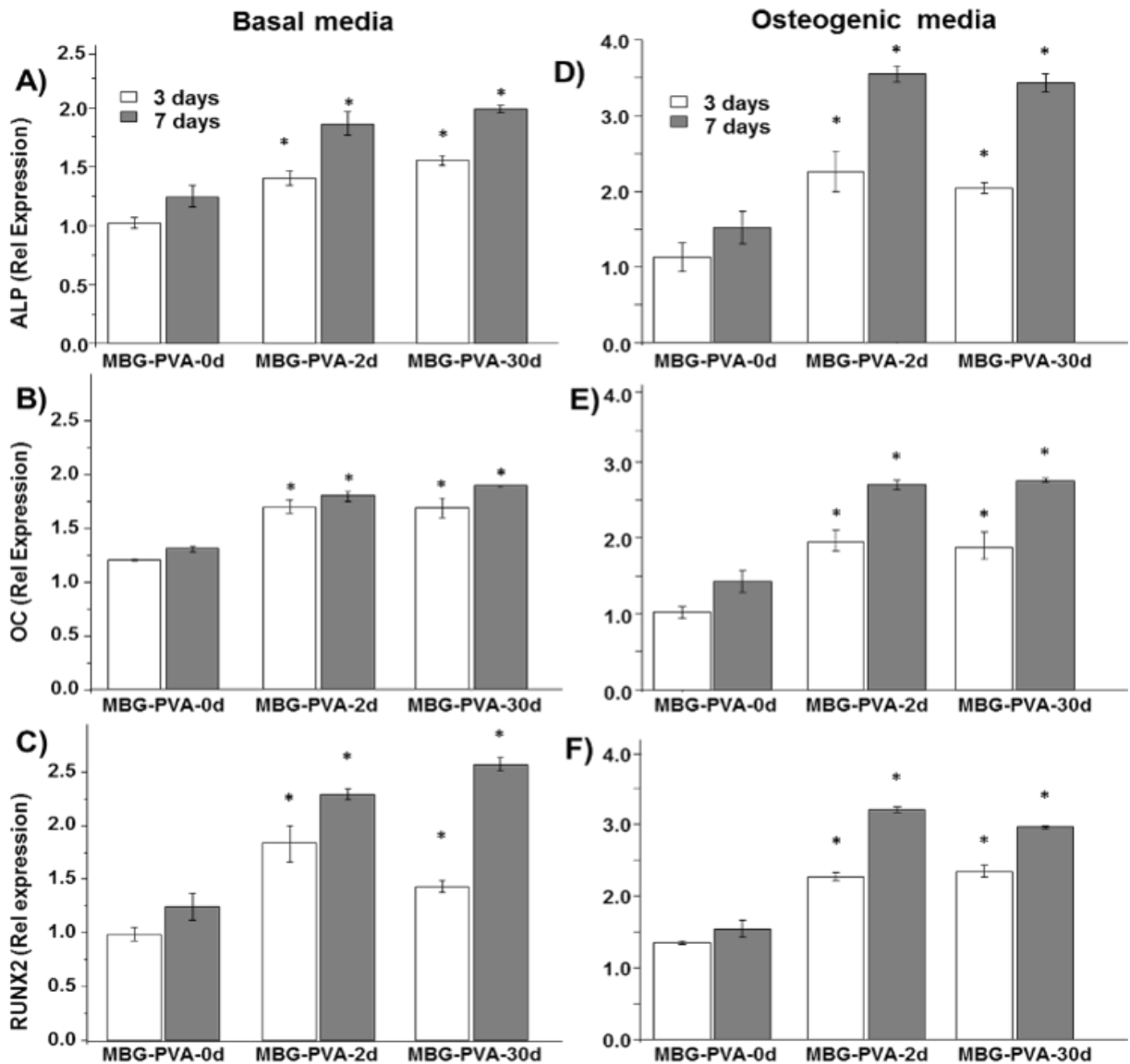

**Figure 7.** MC3T3-E1 relative gene expression to the house keeping gene (GAPDH) at 3 and 7 days of A) ALP B) Osteocalcin C) RunX2 in basal media and D) ALP E) Osteocalcin F) RunX2 in osteogenic media . *<0.05 indicate statistical significance when compared to the MBG-PVA-0d group at the corresponding timepoint.

Figure 8 shows the SEM micrographs and confocal images obtained after 3 and 7 days of culture in basal media for the MBG-PVA-0d, MBG-PVA-2d, and MBG-PVA-30d scaffolds. The SEM images show that the cells spread evenly on the scaffold surface and displayed a morphology typical for cells of an osteoblastic lineage. At 7 days post-seeding, the cells had colonized most of the scaffold and a continuous cell layer was formed in both the MBG-PVA-2d, and MBG-PVA-30d scaffolds demonstrating cell proliferation, which corroborated the Alamar Blue findings. Confocal microscopy images show the complete coverage of the scaffold by cells, confirming the SEM observations.

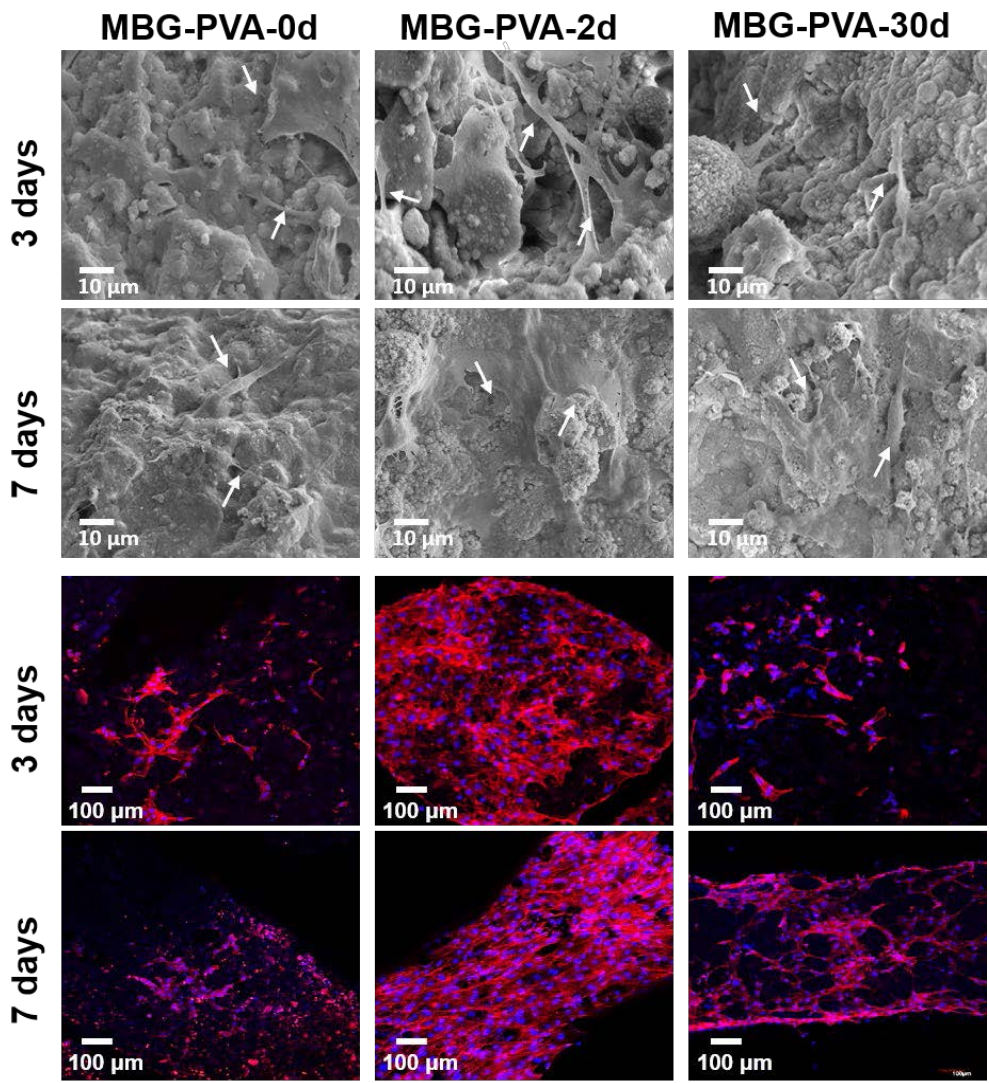

**Figure 8.** SEM and confocal images showing cell morphology on scaffold surfaces after 3 and 7 days in culture (basal media) for the MBG-PVA-0d, MBG-PVA-2d., MBG-PVA-30d. The white arrows show the presence of cells on the scaffold.

This study demonstrated that the immersion of macroporous MBG-PVA scaffolds into PBS is a simple and yet efficient strategy for achieving surface modifications which translated to positively affecting the biological performances of the scaffold. Indeed, these modifications, through biomimetic mineralization, profoundly influenced the scaffold bioactivity [52]. The groups immersed in PBS (MBG-PVA-2d and 30d) showed a significant enhancement of osteogenic related genes when compared to the as-printed group. The improvement of the osteogenic potential of PBS-immersed 3D-printed scaffolds could be topographically-driven via direct cell contact to the MBGs particles and/or to the newly formed apatite layer, which is in agreement with previous reports albeit for different types of topographical modifications [53,54]. The chemical or crystallographic composition of the apatite layer did not influence the osteogenic potential of the resulting scaffold since no differences were observed between MBG-PVA-2d and 30d (where the apatite phase material had matured and become crystalline). Regardless of the crystallographic state of the mineralized layer on the scaffold surface, the formation of this apatite phase has previously been claimed to also occur under in vivo conditions and is an essential step during bioglasses osseo-integration [55,56]. Therefore, the pre-mineralization during PBS immersion may accelerate bone regeneration.

**Conclusions**

MBG-PVA macroporous and fully interconnected scaffolds were fabricated by AM, using PVA as binder to facilitate the printing process. PVA formed a layer over most of the scaffold surface masking the textural properties of the mesoporous ceramic particles. The immersion of MBG-PVA scaffolds in PBS led to the removal of the PVA over the entire scaffold and fostered the nucleation and growth of a newly formed apatite-like phase. This apatite coating reinforced the mechanical properties of the scaffolds insofar it substituted the PVA layer. The chemical, structural and topographical modifications of the scaffold's surface resulted in improved metabolic activity, differentiation and gene expression of pre-osteblast like cells.

ACKNOWLEDGMENT

MNGC acknowledges MINECO Ayudas a la movilidad predoctoral para la realización de estancias breves en centros de I+D. M.V.R. acknowledges funding from the European ResearchCouncil (Advanced Grant VERDI; ERC-2015-AdG Proposal No. 694160). N.G.C. is greatly indebted to Ministerio de Ciencia e Innovación for her predoctoral fellowship and her Ayudas a la movilidad predoctoral para la realización de estancias breves en centros de I+D. The authors also thank to Spanish MINECO (MAT2016-75611-R AEI/FEDER, UE). The authors wish to thank the ICTS Centro Nacional de Microscopia Electrónica (Spain), CAI X-ray Diffraction, CAI NMR, CAI Flow Cytometry and Fluorescence Microscopy of the Universidad Complutense de Madrid (Spain) for their technical assistance. The authors would like to thank Professor Dietmar W. Hutmacher for his highly valuable support and assistance with this paper.

**Conflicts of interest**

There are no conflicts to declare.